\documentstyle[12pt]{article}
\begin{document}
\newcommand{\bqa}{\begin{eqnarray}}
\newcommand{\eqa}{\end{eqnarray}}
\newcommand{\nl}{\nonumber \\}
\newcommand{\eqn}[1]{eq. (\ref{#1})}
\newcommand{\eqs}[1]{eqs. (\ref{#1})}
\title{
\begin{flushright}
\normalsize
\vspace{-4cm}
 CERN-TH/97-372 \\
 hep-ph/9712418 
\end{flushright}
\vspace{4cm}
A simple method for multi-leg loop calculations 2: a general algorithm
\author{R.~Pittau\\
        Theoretical Physics Division, CERN\\
        CH-1211 Geneva 23, Switzerland}
}

\date{}
\maketitle
\thispagestyle{empty}

\vspace{2cm}

\begin{abstract}
The method introduced in a previous paper to simplify the tensorial 
reduction in multi-leg loop calculations is extended to generic 
one-loop integrals, with arbitrary internal masses 
and external momenta.
\end{abstract}

\vspace{4cm}

\begin{flushleft}
CERN-TH/97-372 \\
December 1997  \\
\end{flushleft}

\clearpage
\setcounter{page}{1}
\section{Introduction}
In a previous paper \cite{cpc}, a technique was presented to simplify 
the tensorial reduction of $m$-point one-loop diagrams of the type
\bqa
{\cal M}(p_1,\cdots,p_r;k_1, \cdots , k_{m-1}) ~~=~~
\sum_a\int\,d^n q 
\frac{{\rm Tr}^{(a)}\,[\rlap{/}q \cdots 
\rlap{/}q \cdots\,]}{D_1\,\cdots D_m} \,,
\label{eqgen}
\eqa 
where $p_{1 \cdots\, r}$ are the external momenta of the diagram, 
$k_{1 \cdots\, m-1}$ the momenta in the loop denominators,
defined as
\bqa
D_i ~~=~~ (q+s_{i-1})^2 -m^2_i\,,~~~~~~ 
s_i ~~=~~ \sum_{j=\,0}^{i} k_j ~~~~~~(k_0= 0)\,,
\label{eq0}
\eqa
and ${\rm Tr}^{(a)}$ traces over $\gamma$ matrices, which may contain 
an arbitrary number of $\rlap{/} q$'s.

It was shown that, by assuming at least two massless 
momenta in the set $k_{1\cdots \,m-1}$, 
the traces in \eqn{eqgen} can be rewritten in terms of the
denominators appearing in the diagram, therefore simplifying 
the calculation.

Starting from $m$-point rank-$l$ tensor integrals, 
the algorithm gave at most rank-1 $m$-point functions, 
plus $n$-point rank-$p$ tensor integrals with $n < m$ and $p < l$. 

In this paper, I show how to extend this technique 
when the momenta $k_{1\cdots\,m-1}$ are generic.
On the one hand, this allows to apply the method to more
general problems. On the other hand, the reduction
procedure can therefore be iterated in such a way that,
usually, only rank-1 integrals and scalar functions remain at the end.

In the next section, I introduce the algorithm and
in section 3, I apply it to a specific example.
\section{The general algorithm}
The basic idea is simple.
Given two vectors $\ell_1$ and $\ell_2$, one can
`extract' the $q$ dependence from the traces with
the help of the identity
\bqa
 \rlap{/} q ~~=~~ \frac{1}{2\,(\ell_1 \cdot \ell_2)}\,\left[
               2\,(q \cdot \ell_2)\,\rlap{/} \ell_1   
              +2\,(q \cdot \ell_1)\,\rlap{/} \ell_2 
              -\rlap{/} \ell_1 \rlap{/} q \rlap{/} \ell_2
              -\rlap{/} \ell_2 \rlap{/} q \rlap{/} \ell_1 \right]\,.
\label{eqext}
\eqa
By further assuming $\ell_1^{\,2}= \ell_2^{\,2} = 0$, and making use
of the completeness relations for massless spinors, the 
following result is obtained
\bqa
{\rm Tr}[\rlap{/} q\, \Gamma] &=& \frac{1}{2\,(\ell_1 \cdot \ell_2)}
      \left[\,\, 
    2\,(q \cdot \ell_2)\,{\rm Tr}[\rlap{/} \ell_1 \Gamma] \right. \nl
    &-& \left.
   \left\{q \right\}^{+-}_{1~2} \left\{\Gamma \right\}^{+-}_{2~1}
    -\left\{q \right\}^{-+}_{1~2} 
    \left\{\Gamma\right\}^{-+}_{2~1}
     + (\ell_1 \leftrightarrow \ell_2)\,\, \right]\,,
\label{eqden}
\eqa
where $\Gamma$ represents a generic string of $\gamma$
matrices and 
\bqa
 \left\{\ell_1\,\ell_2\,\cdots\,\ell_n\right\}^{+-}_{i~j}
 ~~\equiv~~ 
 \left\{  1    2\,\cdots\,  n\right\}^{+-}_{i~j}
 ~~\equiv~~
 \bar v_+(\ell_i)\,\rlap{/} \ell_1 \,\rlap{/} \ell_2  
  \cdots \,\rlap{/} \ell_n\, u_-(\ell_j)\,.
\label{eq4}
\eqa 
By iteratively applying the above procedure, together with 
the equations \cite{cpc} 
\bqa
\left\{q \right\}^{-+}_{1~2} \left\{q \right\}^{-+}_{2~1}
  &=& 4\,(q \cdot \ell_1)
\,(q \cdot \ell_2) -2\,q^2\,(\ell_1 \cdot \ell_2) \nl
\left\{q \right\}^{-+}_{1~2} \left\{q \right\}^{-+}_{1~2}
  &=& \frac{2}{\{ b \}^{+-}_{1~2}}
\left[ [q^2 (\ell_1 \cdot \ell_2)-2
           (q \cdot \ell_1)(q \cdot \ell_2)]
    \left\{b \right\}^{-+}_{1~2} \right. \nl
  & &\!\!\!\!\!\!\!\!\!\!\!\!\!\!\!\!\!\!\!\!\!+~\left.2\, 
         [(q \cdot \ell_1)(b \cdot \ell_2)
         -(q \cdot b)(\ell_1 \cdot \ell_2)
         +(q \cdot \ell_2)(\ell_1 \cdot b)]
    \left\{q \right\}^{-+}_{1~2}
            \right],
\label{eqext1}
\eqa
only one $\{q\}^{-+}_{1~2}$ (or its complex conjugate
$\{q\}^{-+}_{2~1}$) survives in each term,
and powers of $q^2$, $(q \cdot \ell_1)$, $(q \cdot \ell_2)$ 
and $(q \cdot b)$ factorize out.

The next step is to reconstruct the denominators from the above scalar
products. By choosing, for example, $b = k_3$ one trivially gets
\bqa
 q^2                  &=& D_1    +m_1^2\,,                    \nl
 2 (q \cdot b     )   &=&  D_4-D_3+m_4^2-m_3^2-(k_1+k_2+k_3)^2
                             +(k_1+k_2)^2\,,
\eqa
but $(q \cdot \ell_1)$ and $(q \cdot \ell_2)$ still remain.

In ref. \cite{cpc} the simple case was studied in which the
diagram in \eqn{eqgen} is such that at least two $k$'s 
(say $k_1$ and $k_2$) are massless. A solution to the
problem is then to take $\ell_1= k_1$ and $\ell_2= k_2$:
\bqa
 2 (q \cdot \ell_1) &=&  D_2-D_1+m_2^2-m_1^2\,,    \nl
 2 (q \cdot \ell_2) &=&  D_3-D_2+m_3^2-m_2^2-(k_1+k_2)^2\,.
\label{case0}
\eqa 
If, in the set $k_{1\cdots\, m-1}$,
only one momentum (say $k_1 \equiv \ell_1$) is massless,
a solution can still be found by decomposing
any other massive momentum (say $k_2$) in terms of massless vectors:
\bqa
k_2~~=~~ \ell_2+\alpha\,\ell_1\,.
\eqa 
The requirement that also $\ell_2$ is massless, implies
\bqa
\alpha~~=~~ \frac{k_2^2}{2 (k_1 \cdot k_2)}\,, 
\label{alpheq}
\eqa
and therefore
\bqa
 2 (q \cdot \ell_1) &=& D_2-D_1+m_2^2-m_1^2\,,                  \\
 2 (q \cdot \ell_2) &=& D_3-(1+\alpha)(D_2+m_2^2)+\alpha(D_1+m_1^2) 
                        +m_3^2-(k_1+k_2)^2 \,. \nonumber
\eqa
When there are no massless $k$'s, a basis of massless vectors
can yet be constructed: 
\bqa
k_1 ~~=~~ \ell_1 +\alpha_1 \ell_2\,,\,\,\,\,\,\
k_2 ~~=~~ \ell_2 +\alpha_2 \ell_1\,.
\eqa
In fact, requiring $\ell_1^{\,2}= \ell_2^{\,2}= 0$ gives 
\bqa
\alpha_1  &=&  \frac{(k_1 \cdot k_2) \pm \sqrt{\Delta}}{k_2^2}\,,~~~
\alpha_2 ~~=~~ \frac{(k_1 \cdot k_2) \pm \sqrt{\Delta}}{k_1^2}\,, \nl
\ell_1    &=&  \beta(k_1-\alpha_1 k_2)\,, ~~~~~~\,
\ell_2   ~~=~~ \beta(k_2-\alpha_2 k_1)\,, \nl
\Delta    &=& (k_1 \cdot k_2)^2-k_1^2 k_2^2\,,\,\, ~~
\beta    ~~=~~ \frac{1}{1-\alpha_1 \alpha_2}\,,
\label{eqdelta}
\eqa 
from which one computes
\bqa
\frac{2 (q \cdot \ell_1)}{\beta} &=& (1+\alpha_1)(D_2-k_1^2+m_2^2)
         -(D_1+m_1^2)                                              \nl
        &-& \alpha_1[D_3+m_3^2-(k_1+k_2)^2]   \,,                  \nl
\frac{2 (q \cdot \ell_2)}{\beta} &=& D_3 +\alpha_2(D_1+m_1^2)
         -(1+\alpha_2)(D_2-k_1^2+m_2^2)                            \nl
        &+& m_3^2-(k_1+k_2)^2\,.
\eqa
When the loop integrals have to be evaluated in $n$
dimensions, the substitution $q \to \underline{q} \equiv q + \tilde q$ 
is needed \cite{cpc,bern}, where $q$ lives in 4 dimensions
and $\tilde q$ is the $(n-4)$-dimensional
part of the integration momentum, such that $(q\cdot \tilde{q})= 0$.
The only change in the previous formulas is that
\bqa
 q^2= D_1-\tilde{q}^{\,2}    +m_1^2\,,
\eqa
and the additional integrals, involving powers of $\tilde{q}^{\,2}$,
can be easily handled as shown in ref. \cite{cpc,mahlon}.

Therefore, the described procedure completely solves the problem, for 
arbitrary $k$'s appearing in the denominators of $n$-dimensional
one-loop diagrams.

If, in the original trace, the number $n_q$ of $\rlap /q$'s 
is less than the number $m$ of loop denominators, 
the algorithm can be iterated until rank-1 functions remain, at most. 
If $n_q \ge m$, owing to the
lack of momenta $k$'s to perform the denominator reconstruction,
residual rank-$p$ two-point integrals remain instead, with
$p \le (2+n_q-m)$. However, two-point tensors are much easier to
handle than generic $m$-point tensors, so that the diagram is 
anyhow simplified.

A last remark is in order.
When some $k$'s become collinear, one is faced with the usual problem 
of singularities generated by the tensor reduction
(for an exhaustive study of this topic, see ref. \cite{glover}).
In fact, denominators appear in eqs. (\ref{eqden}) and (\ref{eqext1}), 
which may vanish, and the quantity $\Delta$ 
in \eqn{eqdelta} is nothing but a Gram determinant.
Even if the occurrence of such singularities cannot be completely
avoided, a better control on them is in general possible \cite{cpc}, 
with respect to traditional techniques \cite{pasve}.
In addition, the analytic expressions can be kept rather compact, 
avoiding, at the same time, the appearance of large-rank tensors.
\section{An example}
To illustrate the method, I compute the reduction for the 
following integral with $n_q =2$:
\bqa
I~~=~~\int d^nq \frac{1}{D_1\,\cdots\,D_m}\,
{\rm Tr}[\,\underline{q}\, \Gamma \, \underline{q}\, \Lambda\,]\,,
\eqa
where, to fix the ideas, $\Gamma$ and $\Lambda$ are strings 
containing an odd number of four-dimensional $\gamma$ matrices.
For convenience of notation, I omit to write the slashes in the traces.
 
Since the integration is performed in $n$ dimensions, 
the denominators are given by \eqn{eq0} with the substitution 
$q \to \underline{q}= q +\tilde{q}$. 

When $m \ge 3$, the algorithm reduces $I$ to 
a sum of scalar and rank-1 integrals. In fact,
by splitting $\underline{q}$ in the numerator, one gets 
\bqa
{\rm Tr}[\,\underline{q}\, \Gamma\, \underline{q}\, \Lambda\,]~~=~~
{\rm Tr}[q \Gamma q \Lambda] - \tilde{q}^{\,2} {\rm Tr}[\Gamma \Lambda]\,,
\eqa
and, by applying the formulas in the previous section,
\bqa
{\rm Tr}[q \Gamma q \Lambda] &=& \frac{1}{2 (\ell_1 \cdot \ell_2)}
\left[ 2(q \cdot \ell_1)E(\ell_2) 
     + 2(q \cdot \ell_2)E(\ell_1)
     - q^2 A-2(q \cdot k_3)\,G \right], \nl
A &=& 2\,{\rm Re} \left[
 \{\Lambda \}^{+-}_{1~1}
 \{\Gamma  \}^{+-}_{2~2}
+\{\Lambda \}^{-+}_{2~2}
 \{\Gamma  \}^{-+}_{1~1}
 -C\,\{k_3 \}^{+-}_{2~1} \right]\,, \nl
G &=& 2\,{\rm Re} \left[
  C\,\{q   \}^{+-}_{2~1} \right]\,, \nl 
C &=& \frac{1}{\{k_3 \}^{+-}_{1~2}}
\left[
 \{\Lambda \}^{-+}_{2~1}
 \{\Gamma  \}^{-+}_{2~1}
+\{\Lambda \}^{+-}_{1~2}
 \{\Gamma  \}^{+-}_{1~2}
\right]\,, \nl
E(\ell) &=& {\rm Tr}[\ell \Gamma q \Lambda]
-\frac{1}{2(\ell_1 \cdot \ell_2)}
\left\{
 {\rm Tr}[\ell_2 q \ell_1 \Gamma \ell \Lambda]
+{\rm Tr}[\ell_1 q \ell_2 \Gamma \ell \Lambda] \right.\nl
&-& \left. 2(k_3 \cdot \ell)\,G -(q \cdot \ell)\,A
\right\}\,.
\label{eqex1}
\eqa
The above equations give the final answer:
\bqa
I &=& \frac{1}{2(\ell_1 \cdot \ell_2)}
\int d^n q \frac{1}{D_1\,\cdots\,D_m}
\left\{
(D_1+m_1^2)\left[ E(\beta\alpha_2\ell_1- \beta\ell_2)-A\right] \right. \nl
&+&  (D_2+m_2^2-k_1^2)\,
  E(\beta\ell_2+\beta\alpha_1\ell_2-\beta\ell_1-\beta\alpha_2\ell_1)
 \nl
&+&  \left(D_3+m_3^2-(k_1+k_2)^2\right)\,
\left[E(\beta\ell_1-\beta\alpha_1 \ell_2)+G \right]
 \nl
&-& \left.  \left(D_4+m_4^2-(k_1+k_2+k_3)^2\right)\,G 
    + \tilde{q}^{\,2} \left(A-2(\ell_1 \cdot \ell_2)\,
{\rm Tr}[\Gamma\Lambda]\right)
\right\}\,,\nl
\beta &=&\frac{1}{1-\alpha_1\alpha_2}\,\,.
\label{eqex2}
\eqa
When $k_{1,\,2}^2 \ne 0$, $\ell_{1,\,2}$ and $\alpha_{1,\,2}$ are 
as in \eqn{eqdelta}.

If $k_1^2= 0$ and $k_2^2 \ne 0$, \eqn{eqex2} still holds
with $\alpha_1 = 0$, $\ell_1 = k_1$ and
$\ell_2= k_2-\alpha_2 k_1$, where $\alpha_2= \alpha$ is given in \eqn{alpheq}.

If   $k_{1,\,2}^2   = 0 $, 
then $\alpha_{1,\,2}= 0$ and $\ell_{1,\,2}= k_{1,\,2}$.

When $m=3$, some terms vanish. This implies $C = G = 0$ and
\bqa
E(\ell)~~=~~ {\rm Tr}[\ell \Gamma q \Lambda]
   + A\, \frac{(q \cdot \ell)}{2(\ell_1 \cdot \ell_2)}\,.
\eqa
\section{Summary}
In this paper, I extended the technique introduced in ref. \cite{cpc} 
to reduce the tensorial complexity of the diagrams
appearing in multi-leg loop calculations.

The method is now applicable to generic one-loop integrals,
with arbitrary internal masses and external momenta.

The algorithm can usually be iterated in such a way that only scalar
and rank-1 functions appear at the end of the reduction.
At worst, higher-rank two-point tensors survive,
independently from the initial number of denominators.

\end{document}